\documentclass[aps,prl, reprint, 
showpacs,showkeys,superscriptaddress,preprintnumbers,nofootinbib]{revtex4-1} 

\usepackage{amsmath,amssymb,bm,mathrsfs}
\usepackage{hyperref}
\usepackage{url}
\usepackage{graphicx}
\usepackage[utf8]{inputenc}

\newcommand\apss{Astrophys. Space Sci.,\,}
\newcommand{\aap}{Astron. Astrophys.\,}
\newcommand{\mnras}{Mon. Not. Roy. Astron. Soc.\,}

\newcommand\unit[1]{\,\mathrm{#1}}
\newcommand\Order{\mathop{\mathcal{O}}}

\begin{document}

\preprint{\tt UT-19-10, IPMU 19-0072}

\title{Dark Matter Heating vs. Rotochemical Heating in Old Neutron Stars} 
\author{Koichi Hamaguchi}
\email{hama@hep-th.phys.s.u-tokyo.ac.jp}
\affiliation{Department of Physics, University of Tokyo, 
Tokyo 113--0033,
Japan}
\affiliation{
Kavli IPMU (WPI), UTIAS, The University of Tokyo, Kashiwa, Chiba 277--8583, Japan
}

\author{Natsumi Nagata}
\email{natsumi@hep-th.phys.s.u-tokyo.ac.jp}
\affiliation{Department of Physics, University of Tokyo, 
Tokyo 113--0033,
Japan}

\author{Keisuke Yanagi}
\email{yanagi@hep-th.phys.s.u-tokyo.ac.jp}
\affiliation{Department of Physics, University of Tokyo, 
Tokyo 113--0033,
Japan}

\begin{abstract}

 Dark matter (DM) particles in the Universe accumulate in neutron stars
 (NSs) through their interactions with ordinary matter. It has been
 known that their annihilation inside the NS
 core causes late-time heating, with which the surface temperature
 becomes a constant value of $T_s \simeq (2-3) \times 10^3$~K for the
 NS age $t \gtrsim 10^{6-7}$~years. This conclusion is,
 however, drawn based on the assumption that the beta equilibrium is
 maintained in NSs throughout their life, which turns out to
 be invalid for rotating pulsars. The slowdown in the pulsar rotation
 drives the NS matter out of beta equilibrium, and the
 resultant imbalance in chemical potentials induces late-time heating,
 dubbed as \textit{rotochemical heating}. This effect can heat a NS   
 up to $T_s \simeq 10^6$~K for $t \simeq 10^{6-7}$~years. In fact,
 recent observations found several old NSs whose surface
 temperature is much higher than the prediction of the standard cooling
 scenario and is consistent with the rotochemical heating. Motivated by these
 observations, in this letter, we reevaluate the significance of the
 DM heating in NSs, including the effect
 of the rotochemical heating. We then show that the signature of
 DM heating can still be detected in old ordinary pulsars, while it is
 concealed by the rotochemical heating for old millisecond
 pulsars. To confirm the evidence for the DM heating, however,
 it is necessary to improve our knowledge on nucleon pairing gaps as
 well as to evaluate the initial period of the pulsars accurately. In any cases, a
 discovery of a very cold NS can give a robust constraint on the
 DM heating, and thus on DM models. To demonstrate
 this, as an example, we also discuss the case that the DM is the 
 neutral component of an electroweak multiplet,
 and show that an observation of a NS with $T_s \lesssim 10^3$~K imposes a
 stringent constraint on such a DM candidate.

\end{abstract}

\maketitle

\section{Introduction}

Despite the firm evidences for dark matter (DM), its nature has not been
unraveled yet. A promising possibility is that DM consists of a weakly-interacting
massive particle (WIMP), as its thermal relic can naturally
explain the observed DM density. WIMPs have interactions with ordinary
matter other than the gravitational interactions, through which we may
detect their signature. There have been various proposals for such experiments so
far, \textit{e.g.}, the DM direct/indirect searches and the direct production of
WIMPs at colliders.

Among other things, the detection of DM signature
through the observation of neutron star (NS) surface temperature offers
a distinct strategy for testing DM models
\cite{Kouvaris:2007ay, Bertone:2007ae, Kouvaris:2010vv,
deLavallaz:2010wp}. DM particles which are trapped by the gravitational
potential of a NS accumulate in the core after they have
lost their kinetic energy through the scattering with the NS matter. 
These DM particles eventually annihilate and heat the NS. 
At late times, this heating effect balances with the energy loss due to the photon
emission from the NS surface, and its surface temperature $T_s$ is kept
constant at $T_s \simeq 2 \times 10^3$~K. This consequence is in
stark contrast to the prediction in the standard NS cooling theory
\cite{Yakovlev:1999sk, Yakovlev:2000jp, Yakovlev:2004iq}, where isolated
NSs cool down to $T_s < 10^3$~K for the NS age $t \gtrsim 5 \times
10^6$~years. This implies that we can in principle test this DM heating
scenario by measuring the surface temperature of old NSs. See
Refs.~\cite{Bramante:2017xlb, Baryakhtar:2017dbj, Raj:2017wrv,
Chen:2018ohx, Bell:2018pkk, Camargo:2019wou, Bell:2019pyc} for recent
studies on the DM heating.

In the previous studies of the DM heating, as well as in the standard NS 
cooling scenario, it is assumed that nucleons and charged leptons in
NSs are in beta equilibrium throughout the time evolution of the NSs. For actual
rotating NSs, however, this assumption is found to be invalid. A
constant decrease in the rotational rate of a NS leads to a
continuous reduction in the centrifugal force, and thus the NS keeps
being contracted. This then changes the chemical equilibrium condition
among nucleons and leptons all the time
\cite{Reisenegger:1994be}. On the other hand, the rates of the beta
processes are highly suppressed at late times, and thus the NS matter
is unable to follow the change in the chemical equilibrium condition. 
As a result, the imbalance in the chemical potentials of nucleons and
leptons increases constantly, which is to be partially dissipated as
heat~\cite{Reisenegger:1994be, 1992A&A...262..131H,
1993A&A...271..187G}. This heating effect, called the rotochemical
heating, can raise the NS surface temperature up to $T_s
\simeq 10^6$~K for $t \simeq 10^{6-7}$~years \cite{Fernandez:2005cg,
Villain:2005ns, Petrovich:2009yh, Pi:2009eq, Gonzalez-Jimenez:2014iia,
Yanagi:2019vrr}. 

In fact, recent observations of old NSs suggest that
such a heating mechanism indeed operates. For instance, the surface temperature
of the millisecond pulsar (MSP) J0437-4715, whose age is estimated to be
$t\simeq (6-7) \times 10^{9}\unit{yr}$, is measured to be $T_s^\infty \sim
3 \times 10^5\unit{K}$ \cite{Kargaltsev:2003eb, Durant:2011je, Gonzalez-Caniulef:2019wzi},
where $T_s^\infty$ represents the red-shifted temperature  at
the infinite distance. Other examples of NSs whose surface temperature
is higher than the prediction in the standard cooling theory include 
the MSP J2124-3358~\cite{Rangelov:2016syg} and ordinary pulsars
J0108-1431~\cite{Mignani:2008jr} and B0950+08~\cite{Pavlov:2017eeu}. 
On the other hand, for J2144-3933, there is only
an upper bound on the surface temperature: $T_s^\infty < 4.2 \times
10^4\unit{K}$ \cite{Guillot:2019ugf}. Quite interestingly, it is shown
in Ref.~\cite{Yanagi:2019vrr} that all of these observations can be
explained by the effect of the non-equilibrium beta processes in a
consistent manner. 

It should be emphasized that the non-equilibrium beta effect
mentioned above is an inevitable consequence of a rotating pulsar, not
an ad-hoc assumption to invent a heating mechanism. Given this heating
mechanism intrinsic to actual NSs, can we still expect to detect the
signature of the DM heating in old NSs? This is the question we address
in this letter.

\section{Minimal cooling}

We first review the minimal cooling scenario \cite{Page:2004fy,
Gusakov:2004se, Page:2009fu}, which is the basis 
of the following discussions. In this paradigm, it is assumed that a NS is
comprised of nucleons, electrons, and muons and they are in
beta equilibrium. NSs cool via the emission of neutrinos from
the core and photons from the surface. The photon emission dominates the
neutrino emission at late times, $t \gtrsim 10^5$~years. It is found
\cite{Yakovlev:1999sk, Yakovlev:2000jp, Yakovlev:2004iq} that the
thermal relaxation in a NS is completed in $t \lesssim 10^{2}$~years,
and after that the red-shifted internal temperature defined by 
$T^\infty \equiv T(r) e^{\Phi(r)}$ becomes constant in the core,
where $T(r)$ denotes the local temperature at the distance $r$ from the
center and  $e^{2\Phi(r)}=-g_{tt}(r)$ is the time component of the
metric at the position. The time evolution of this red-shifted
temperature is then determined by 
\begin{align}
  \label{eq:time-evl}
  C\frac{dT^\infty}{dt} &= -L_\nu^\infty -L_\gamma^\infty + L_H^\infty,
\end{align}
where $L_\nu^\infty$ and $L_\gamma^\infty$ are the red-shifted
luminosities of the neutrino and photon emissions, respectively, and $C$
is the total heat capacity of the NS. $L_H^\infty$
represents the heating power, which vanishes
in the minimal cooling. 

The photon emission luminosity is given by $L_\gamma=4\pi R^2\sigma_B
T_s^4$, where $\sigma_B$ is the Stefan-Boltzmann constant and $R$ is the NS radius. The surface
temperature $T_s$ differs from the
internal temperature due to the shielding effect of the envelope. To
relate these temperatures, we use the formula given in
Ref.~\cite{Potekhin:1997mn} for $T \gtrsim 10^4$~K. This relation
depends on the amount of light elements in the envelope, which is
parametrized by their total mass $\Delta M$. The relation in
Ref.~\cite{Potekhin:1997mn} cannot be used for $T
\lesssim 10^4$~K, for which we instead use the relation for the heavy-element
envelope given in Ref.~\cite{1983ApJ...272..286G}.

There are multiple processes for the neutrino emission. Among them, the modified Urca
process \cite{1995A&A...297..717Y, Gusakov:2002hh} and the pair
breaking and formation (PBF) process~\cite{1976ApJ...205..541F, Voskresensky:1987hm,
Senatorov:1987aa, Yakovlev:1998wr, Kaminker:1999ez, Leinson:2006gf} are the dominant processes in the minimal cooling. 
The modified Urca process consists of the reactions 
\begin{align}
  &n + N \to p + N + \ell + \bar\nu_\ell \,,
   \notag\\
  & p + N + \ell \to n + N + \nu_\ell\,,
    \label{eq:murca}
\end{align}
where
$N =n$ or $p$ and $\ell =
e, \mu$. The rate of this process goes as $\propto T^8$ at high
temperatures, but is highly suppressed after the onset of nucleon
superfluidity. In the NS core, protons form singlet pairings
and neutrons form triplet pairings once the temperature becomes lower
than the corresponding critical temperatures \cite{Page:2013hxa,
Haskell:2017lkl, Sedrakian:2018ydt}. These nucleon parings yield an
energy gap in the spectrum of nucleon quasi-particles, which gives a
Boltzmann suppression factor to the reaction rates. The pairing gaps also
suppress the contribution of nucleons to the heat capacity in
Eq.~\eqref{eq:time-evl} at low temperatures \cite{Yakovlev:1999sk}. 

The PBF process, on the other hand, operates only after the onset of
nucleon superfluidity. This reaction proceeds along with the breaking
of a nucleon pairing due to thermal fluctuation and its successive
reformation, during which neutrinos are emitted. Since it
requires the formation of nucleon pairings, it can occur only for $T <
T_C^{(N)}$, where $T_C^{(N)}$ is the critical temperature of
nucleon superfluidity. For $T \ll T_C^{(N)}$, the PBF process also 
suffers from the Boltzmann suppression due to the energy gap in the
nucleon spectrum.

In summary, in the minimal cooling scenario, a NS cools via the
emission of photons and neutrinos. For the neutrino emission, we
consider the modified Urca and PBF processes, and neglect other
subdominant processes for brevity.
Notice that the fast cooling processes such as the direct Urca process
are not included in the minimal cooling. It is known that the direct Urca
process can occur only in heavy NSs~\cite{Lattimer:1991ib}; for
instance, in the case of the Akmal-Pandharipande-Ravenhall (APR)
equation of state (EOS)~\cite{Akmal:1998cf}, which we use in the
following analysis, the direct Urca can occur for $M\gtrsim
1.97\,M_{\odot}$, where $M$ and $M_{\odot}$ are the NS mass and the solar mass, respectively. In this work, we follow the minimal cooling paradigm
and do not consider the fast cooling processes.

\section{Rotochemical heating}

In the minimal cooling,
nucleons and leptons in NSs are assumed to be in beta
equilibrium. On the other hand, as discussed in the introduction, the
local chemical-equilibrium condition in an actual NS changes continuously
because of the constant reduction in the centrifugal force, which is
caused by the slowdown in the NS rotation. It turns out that the NS
system cannot follow this change at late times since the modified Urca
process is strongly suppressed at low temperatures. Therefore, it is
necessary to take account of the out-of-beta-equilibrium effect
for the discussion of old NSs. 

The deviation from the beta equilibrium is quantified by an imbalance
in the chemical potentials of nucleons and leptons: $\eta_\ell \equiv \mu_n -
\mu_p - \mu_\ell$. As discussed in 
Ref.~\cite{Reisenegger:1996ir}, the diffusion timescale of the
chemical imbalance is short enough so that we can regard the red-shifted
imbalance parameters $\eta_\ell^\infty \equiv \eta_\ell e^{\Phi(r)}$ as
constant throughout the NS core. The time evolution of $\eta_\ell^\infty$ is
then obtained by solving a couple of differential equations given in
Ref.~\cite{Fernandez:2005cg}, to which two classes of effects contribute
competitively. One is the terms proportional to the difference between
the reaction rates of the processes \eqref{eq:murca}, $\Delta
\Gamma_{M,N\ell}$,  which reduce
$|\eta_\ell^\infty|$, \textit{i.e.}, restore the system back to beta
equilibrium. The other terms are proportional to $\Omega \dot{\Omega}$,
where $\Omega$ is the angular velocity of the NS, which increase the
value of $\eta_\ell^\infty$ as the NS is slowing down and thus drive
the system out of beta equilibrium. Once the latter contribution
dominates the former, $\eta_\ell^\infty$ keeps increasing and the
system deviates far from beta equilibrium. 

The energy stored in the chemical imbalance is partially released as
heat. The heating rate per unit volume is
given by $\eta_\ell \cdot \Delta \Gamma_{M,N\ell}$, and $L_H^\infty$ in
Eq.~\eqref{eq:time-evl} is
obtained by integrating this quantity (with a red-shift factor) over the
NS core. $\Delta \Gamma_{M,N\ell}$ is computed in the 
literature \cite{Fernandez:2005cg, Villain:2005ns, Petrovich:2009yh,
Pi:2009eq, Gonzalez-Jimenez:2014iia, Yanagi:2019vrr}, which is again
suppressed in the presence of nucleon superfluidity due to the energy
gap. A distinct feature of the non-equilibrium beta process is that
it is strongly enhanced once $\eta_\ell$ exceeds a certain threshold
value, $\Delta_{\mathrm{th}} = \mathrm{min}\{3\Delta_n +
\Delta_p, \Delta_n + 3\Delta_p\}$ \cite{Reisenegger:1996ir}, where
$\Delta_p$ and $\Delta_n$ are the proton singlet and neutron
triplet\footnote{We use the $^3P_2$ ($m_J = 0$) gap for neutron, where
the gap is given by $\delta_n = \Delta_n \sqrt{1+ 3 \cos^2 \theta}$ with
$\theta$ the angle between the neutron momentum and the quantization
axis. } gaps, respectively. At early times, $\eta_\ell$ is negligibly
small and thus the heating is ineffective. At later times,
$\eta_\ell^\infty$ monotonically increases due to the spin-down of the
NS, and once it exceeds the threshold $\Delta_{\mathrm{th}}$, $\Delta
\Gamma_{M,N\ell}$ is strongly enhanced and the rotochemical heating
becomes effective.

The increase rate of $\eta_\ell^\infty$ depends on
the slowdown factor $\Omega \dot{\Omega}$. 
We assume that the spin-down rate of NSs is given by the
energy loss due to the magnetic dipole radiation and thus follows the
power-law deceleration: $\dot\Omega(t) = -k\Omega(t)^3$ with $k$ a
positive constant. By solving this equation, we obtain $\Omega (t) =
2\pi/\sqrt{P_0^2 + 2 P\dot{P} \,t}$, where $P$ and $\dot P$ are
the rotational period of the NS and its time derivative at the time $t$,
respectively, $P_0$ denotes the initial period, and $k =
P\dot{P}/(4\pi^2)$. $P\dot{P}$ is related
to the surface magnetic field by $B_s \simeq 3.2 \times
10^{19}\, (P\dot P/\mathrm{s})^{1/2}\unit{G}$ for a NS of radius $R =
10\unit{km}$ and moment of inertia $I=10^{45}\unit{g}\unit{cm^2}$. These
expressions show that for a larger $P_0$, $|\Omega
\dot{\Omega}|$ becomes smaller, which results in a suppression in
the increase rate of $\eta_\ell^\infty$.

\begin{figure}
{\includegraphics[clip, width = 0.45 \textwidth]{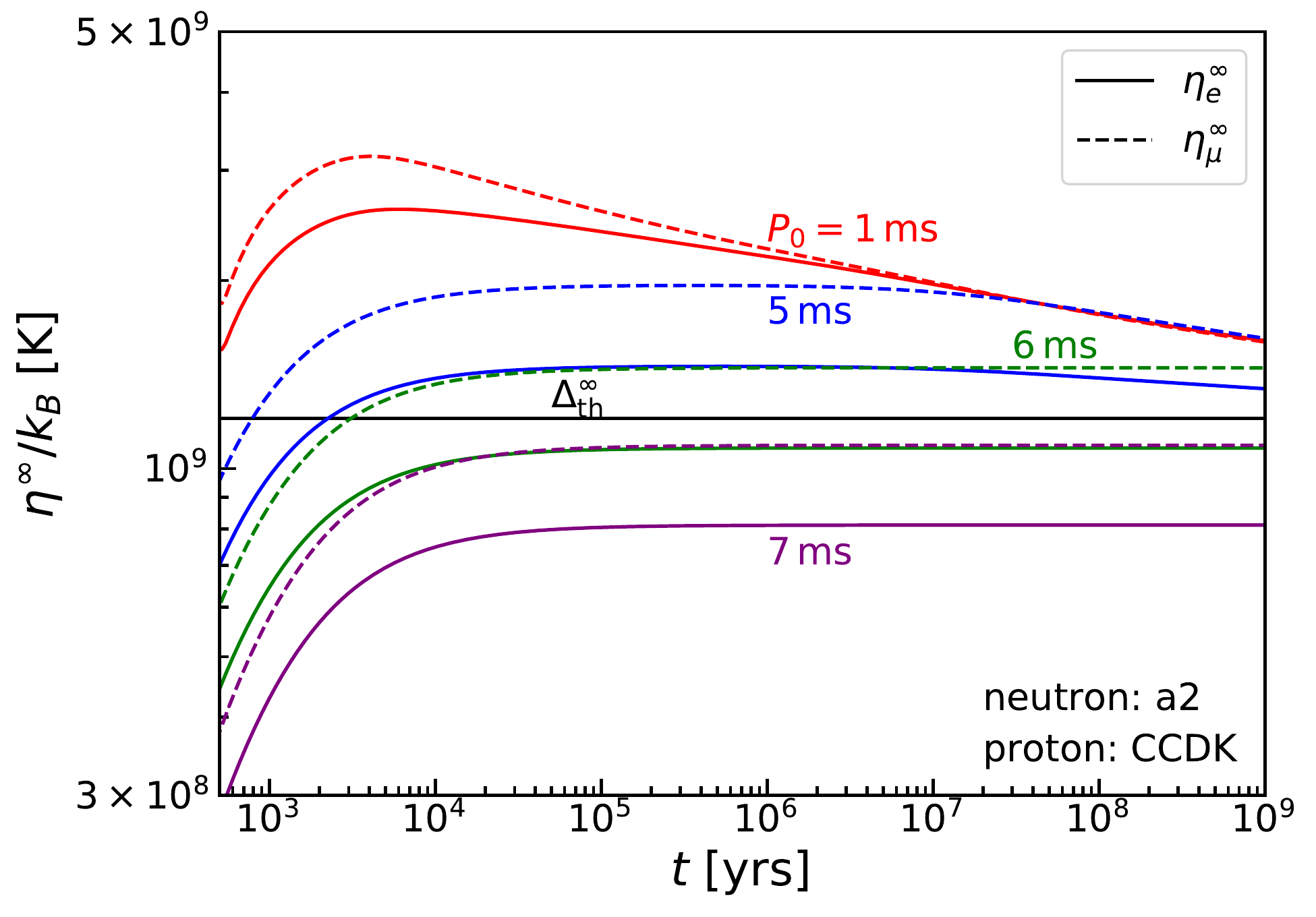}} 
\caption{
The time evolution of $\eta_\ell^\infty$ for different values of
 $P_0$. The horizontal line shows
 $\Delta_{\mathrm{th}}^\infty$ defined in the text. We use $M = 1.4 M_\odot$, $P = 1$~s,
 $\dot{P} = 1 \times 10^{-15}$, and $\Delta M/M = 1 \times 10^{-15}$.  
}
\label{fig:eta_P0}
\end{figure}

To see this in more detail, in Fig.~\ref{fig:eta_P0}, we show
the time evolution of $\eta_\ell^\infty$ for different values of the
initial period $P_0$. Throughout this work, we follow the
analysis given in Ref.~\cite{Yanagi:2019vrr} for the calculation of the
non-equilibrium beta processes. The horizontal line indicates the
minimal value of $\Delta_{\mathrm{th}}$ in the NS core multiplied by a red-shift
factor at the position, which we denote by $\Delta_{\mathrm{th}}^\infty$.  
For nucleon pairing
gaps, we use the CCDK model \cite{Chen:1993bam} for the proton singlet
gap and the ``a2'' model \cite{Page:2013hxa} for the neutron triplet
gap, which is used in Ref.~\cite{Page:2010aw}.\footnote{This choice of the
pairing gaps is found to be compatible with the observed rapid cooling
of the NS in Cassiopeia A \cite{Page:2010aw,
2011MNRAS.412L.108S, Ho:2014pta, Hamaguchi:2018oqw}. } 
This figure shows that as $P_0$ gets larger the increase rate of
$\eta_\ell$ gets smaller. It is also found that for $P_0 > 7$~ms,
$\eta_\ell^\infty$ never exceeds the rotochemical threshold; in this
case, we expect that the rotochemical heating is fairly suppressed, as
we will confirm in the following analysis.

\section{Dark matter heating}

The accretion and annihilation of WIMP DM in a NS can be another heating
source \cite{Kouvaris:2007ay, Bertone:2007ae, Kouvaris:2010vv,
deLavallaz:2010wp}. DM falling toward a NS hits the NS if its impact
parameter is smaller than $b_{\mathrm{max}} = R
(v_{\mathrm{esc}}/v_{\mathrm{DM}}) e^{-\Phi(R)}$
\cite{Goldman:1989nd}, where $G$ is the
gravitational constant, $v_{\mathrm{DM}}$ is the DM velocity
distant from the NS, and
$v_{\mathrm{esc}} = (2GM/R)^{1/2}$ is the escape velocity. 
The rate of DM particles hitting the NS is then obtained as $\dot{N} \simeq \pi b_{\mathrm{max}}^2
v_{\mathrm{DM}} (\rho_{\mathrm{DM}}/m_{\mathrm{DM}})$, where $\rho_{\mathrm{DM}}$ 
and $m_{\mathrm{DM}}$ are the local energy density and mass of DM, respectively. 
We use $\rho_{\mathrm{DM}} = 0.42~\mathrm{GeV} \cdot \mathrm{cm}^{-3}$ and $v_{\mathrm{DM}} 
= 230~\mathrm{km}\cdot \mathrm{s}^{-1}$ \cite{Pato:2015dua} in what follows. 
A more accurate expression of $\dot{N}$ is given in
Ref.~\cite{Kouvaris:2007ay}, which we use in the following analysis.

It is found that an electroweak/TeV-scale WIMP DM is captured in NSs after one
scattering if the DM-nucleon scattering cross section is larger
than $\sigma_{\mathrm{crit}} \simeq R^2 m_N/M$, with $m_N$  the nucleon
mass \cite{Kouvaris:2007ay}. After the DM is trapped, the rest of 
its kinetic energy is soon lost by successive scatterings with the NS matter.
DM particles then accumulate in the NS core and eventually annihilate.
As shown in Ref.~\cite{Kouvaris:2010vv}, for a typical WIMP, its annihilation and
capture rates become in equilibrium in old NSs. As a result, the
contribution of the DM heating to the luminosity $L_H^\infty$ in
Eq.~\eqref{eq:time-evl} is computed as
\begin{equation}
 L_H^\infty|_{\mathrm{DM}} = e^{2\Phi(R)} \dot{N} m_{\mathrm{DM}}\left[
\chi + (\gamma -1)
\right] ~,
\label{eq:lhdm}
\end{equation}
where $\gamma =
1/\sqrt{1-v_{\mathrm{esc}}^2}$ and $\chi$ is the fraction of the
annihilation energy transferred to heat \cite{Kouvaris:2010vv}. In what
follows, we take $\chi = 1$ unless otherwise noted. The first term in Eq.~\eqref{eq:lhdm} represents the
heat from the DM annihilation, while the second term corresponds to the
deposit of the kinetic energy of the incoming WIMP DM
\cite{Baryakhtar:2017dbj}. 

If we neglect the rotochemical heating, the DM heating balances with the
cooling due to the photon emission at late times, \textit{i.e.},
$L_H^\infty|_{\mathrm{DM}}
\simeq L_\gamma^\infty$. This condition fixes the NS surface temperature
to be a few thousand K, which has been regarded as a smoking-gun
signature of the DM heating \cite{Kouvaris:2007ay, Bertone:2007ae, Kouvaris:2010vv,
deLavallaz:2010wp}. Below, we study if this signature can
still be seen even in the presence of the rotochemical heating.

\section{Results}

Now we examine the time evolution of the NS temperature by including
all of the effects discussed above. We first consider a NS which
models a typical ordinary pulsar, where we fix $M = 1.4 M_\odot$,
$P = 1$~s, $\dot{P} = 1 \times 10^{-15}$, and $\Delta M/M = 1 \times
10^{-15}$. The initial values of $T^\infty$
and $\eta_\ell^\infty$ are taken to be $T^\infty = 10^{10}$~K and
$\eta_\ell^\infty = 0$, respectively. We find that the following
results have little dependence on the choice of these parameters.

\begin{figure}
{\includegraphics[clip, width = 0.45 \textwidth]{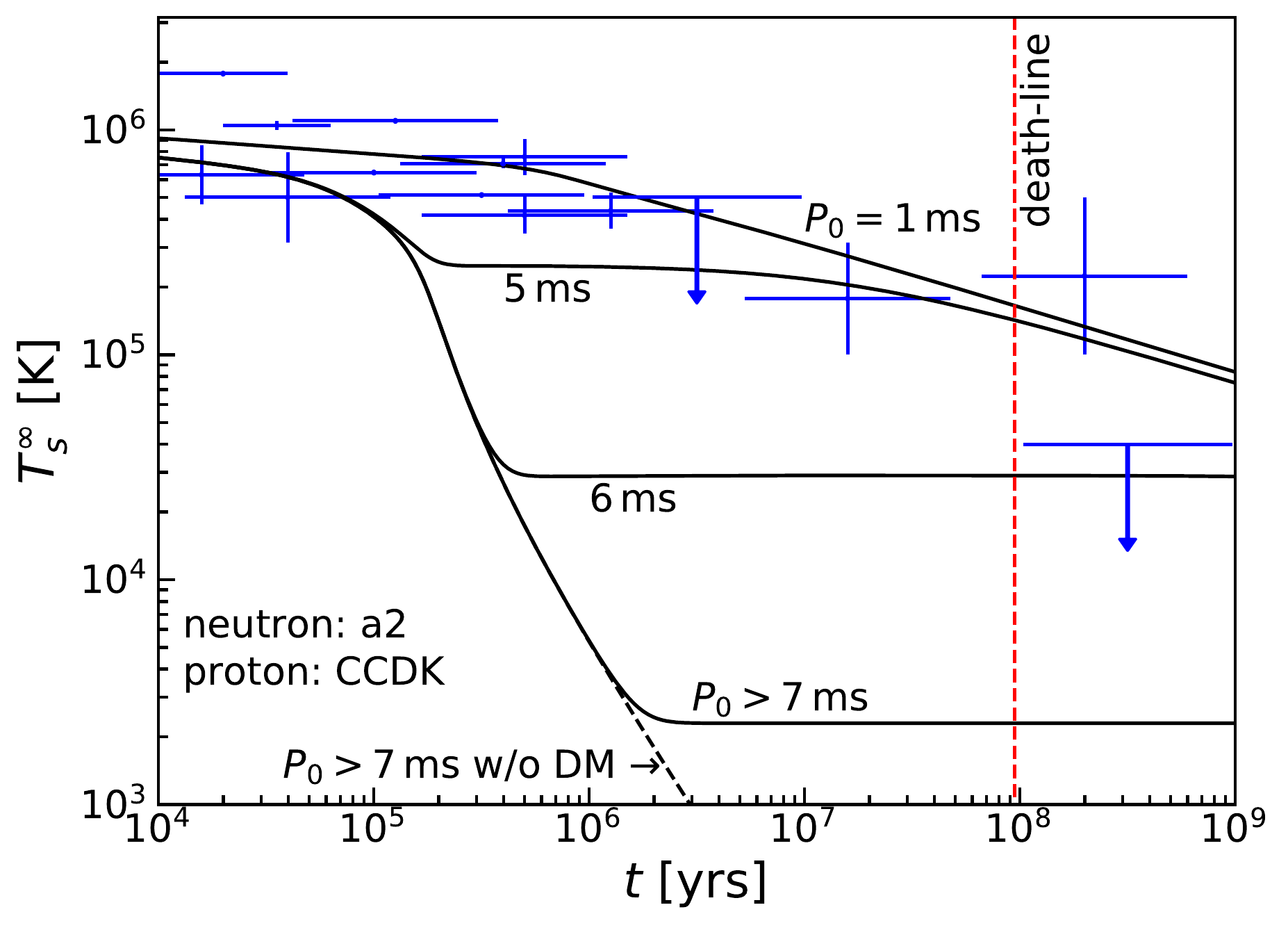}} 
\caption{
The time evolution of $T_s^\infty$ for different values of $P_0$.
For $P_0=1$, 5, and 6~ms, the time evolution with and without DM heating are indistinguishable and the lines overlap. For $P_0>7$~ms, the solid (dashed) line represents the case with (without) DM heating. The blue crosses
 show the temperature data of ordinary pulsars \cite{Yanagi:2019vrr}.
 The down arrows indicate the upper limits on $T_s^\infty$.
 The
 vertical red dashed line shows the death-line \cite{1992A&A...254..198B}. 
}
\label{fig:a2_CCDK_P0}
\end{figure}

In Fig.~\ref{fig:a2_CCDK_P0}, we show the time evolution of $T_s^\infty$
for different values of $P_0$ in the black solid
lines. 
For $P_0=1$, 5, and 6~ms, the time evolutions with and without DM heating are indistinguishable and the lines overlap. For $P_0>7$~ms, the solid (dashed) line represents the case with (without) DM heating.
We again use the CCDK \cite{Chen:1993bam} 
and ``a2'' \cite{Page:2013hxa} models for the proton and neutron 
gaps, respectively. We also show the observed temperatures
of old ordinary pulsars with blue crosses, where the lines indicate the
uncertainties; we take this data from Ref.~\cite{Yanagi:2019vrr}. This
figure shows that for $P_0 = 1$~ms the surface temperature remains as
high as ${\cal O}(10^5)$~K for $t \gtrsim 10^6$~years since the
rotochemical heating is quite effective. The temperature curve in
this case is consistent with most of the observed temperatures, but the DM
heating effect is completely hidden by the rotochemical heating
effect. For a larger $P_0$, $T_s^\infty$ at late times gets lower, and 
for $P_0 > 7$~ms, it becomes independent of the initial period. In this case, the
rotochemical heating is ineffective since $\eta_\ell$ does
not exceed the rotochemical threshold, as we have seen in
Fig.~\ref{fig:eta_P0}. Thus, the late-time temperature is determined by
the DM heating, with $T_s^\infty \simeq 2 \times 10^3$~K. Notice that
a NS cools down to this temperature before it reaches the conventional
death-line \cite{Ruderman:1975ju, 1992A&A...254..198B},\footnote{We however note that
the theoretical estimation of the death-line suffers from
huge uncertainty, and thus one should not take this bound too
seriously. Indeed, as can be seen from
Fig.~\ref{fig:a2_CCDK_P0}, J2144-3933, \textit{e.g.}, is located beyond
the conventional death-line, though its pulsation is detected
\cite{1999Natur.400..848Y}. For more discussions on the death-line, see
Refs.~\cite{1993ApJ...402..264C, Zhang:2000rd, Zhang:2002uh, Zhou:2017hfm}. } 
$B_s/P^2 = 0.17 \times 10^{12}~\mathrm{G} \cdot \mathrm{s}^{-2}$,
shown by the
red dashed line in Fig.~\ref{fig:a2_CCDK_P0}. Therefore, it is
possible to detect the DM heating effect via the temperature observation
of ordinary pulsars if their initial period is sufficiently large.

\begin{figure}
{\includegraphics[clip, width = 0.45 \textwidth]{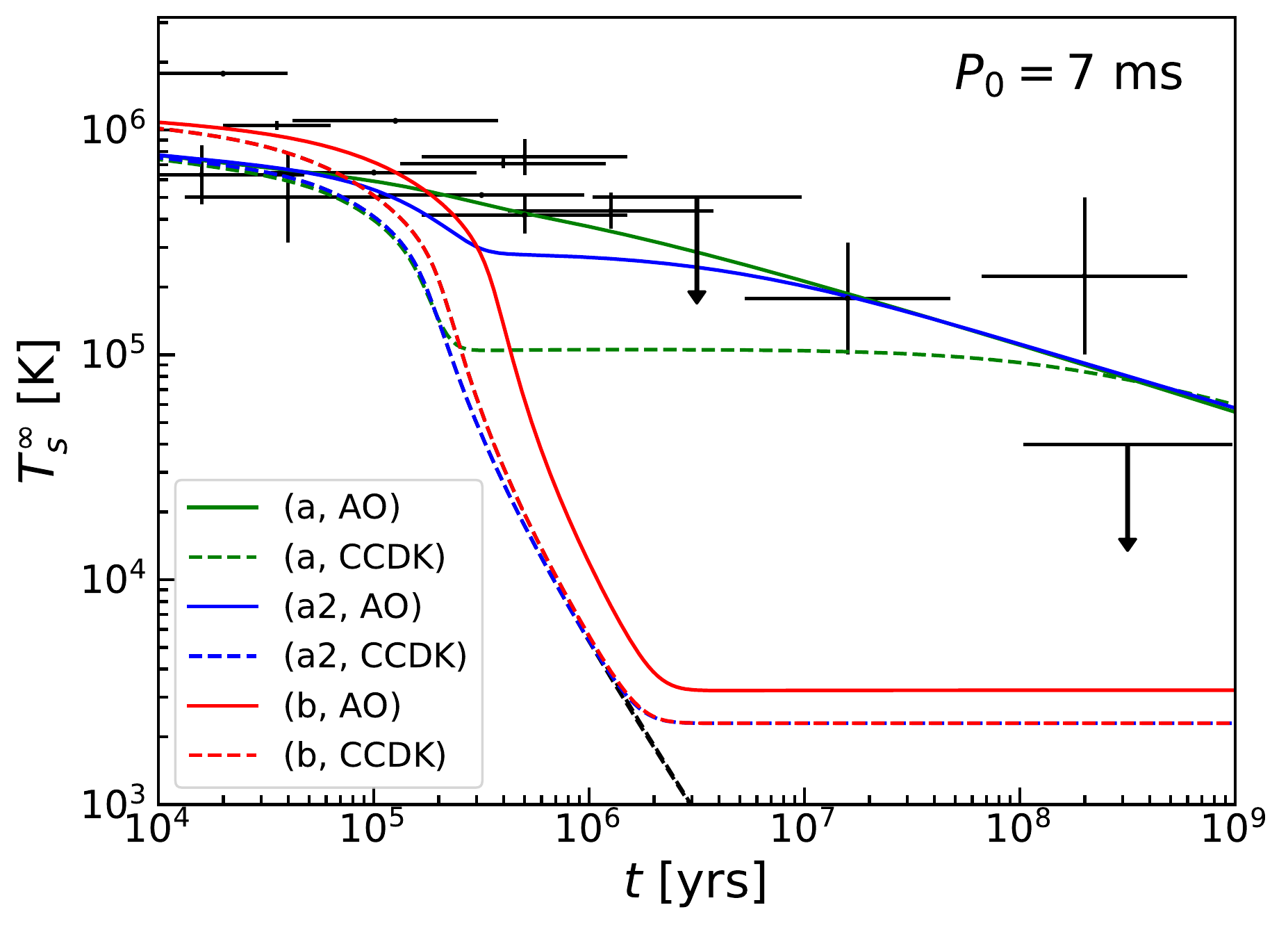}} 
\caption{
The time evolution of $T_s^\infty$ for different nucleon pairing gaps with $P_0 = 7$~ms. 
The green, blue and red lines correspond to 
the ``a'', ``a2'' and ``b'' models for the neutron gap, while the solid and dashed lines show the cases for the AO and CCDK models of the
proton gap, respectively.
For the (a2, CCDK) and (b, CCDK) gap models, the time evolution without DM are also shown in black dashed lines. In the other gap models, the time evolutions with and without DM are almost the same. The black crosses show the temperature data of ordinary pulsars.
}
\label{fig:CP_10ms}
\end{figure}

We note, however, that the lower limit on $P_0$ 
for the condition that the DM heating effect is detectable highly
depends on the nucleon pairing gaps. To see this, in
Fig.~\ref{fig:CP_10ms}, we show the time evolution of $T_s^\infty$ for
different choices of nucleon pairing gaps, with $P_0 = 7$~ms.  
The blue lines correspond to 
the ``a2'' model \cite{Page:2013hxa}, while the green and red lines are
for the ``a'' and ``b'' models in Ref.~\cite{Page:2004fy} for the
neutron gap. The solid and dashed lines show the cases for the
AO \cite{Amundsen:1984qq} and CCDK \cite{Chen:1993bam} models of the
proton gap, respectively. As we see, for the AO model, whose gap
is smaller than that of the CCDK model, the late-time temperature is
predicted to be higher than $T_s^\infty \simeq 2 \times 10^3$~K; in this
case, $\Delta_{\mathrm{th}}^\infty$ is rather small, and thus $\eta_\ell$ can overcome
the rotochemical threshold at late times even for $P_0 = 7$~ms, making
the rotochemical heating operative. On the other hand, for the CCDK proton and "a2" or "b" neutron pairings, the rotochemical heating is
ineffective and thus we can see the DM heating effect at late times.
The results shown in Fig.~\ref{fig:CP_10ms} demonstrate that it is
crucial to take account of both proton and neutron pairings in
order to evaluate the effect of non-equilibrium beta processes
appropriately \cite{Yanagi:2019vrr}. 

We also found that the rotochemical heating does not operate for
any choice of pairing gaps if the initial period is as large as
100~ms\footnote{Since the pairing gap has density dependence, the critical value of $P_0$, above which the rotochemical heating is ineffective, depends also on other star parameters such as NS mass. We find the critical $P_0$ is at most $\Order(10)\unit{ms}$.}---in this case, the late-time surface temperature is always
determined by the DM heating. It is intriguing that recent
studies suggest that the initial period distribution extends to well beyond 100~ms; it is independently estimated from the kinematic age of several tens of observed NSs~\cite{2012Ap&SS.341..457P, 2013MNRAS.430.2281N,  Igoshev:2013rqf}, the population synthesis of pulsars~\cite{FaucherGiguere:2005ny, Popov:2009jn, Gullon:2014dva, Gullon:2015zca}, or the supernova simulation for proto-NSs~\cite{Muller:2018utr}. Hence, we
expect that there are quite a few ordinary pulsars that can be a probe
of the DM heating in future observations.

Finally, let us consider MSPs. 
In this case, $|\Omega \dot{\Omega}|$ is
much larger than that for ordinary pulsars, and thus $\eta_\ell$ 
can always exceed the rotochemical threshold at late
times. Therefore, the rotochemical heating is highly effective for MSPs
\cite{Yanagi:2019vrr}. Although this feature is advantageous for explaining the
old warm MSPs such as J0437-4715 \cite{Kargaltsev:2003eb,
Durant:2011je, Gonzalez-Caniulef:2019wzi} and J2124-3358
\cite{Rangelov:2016syg}, this makes MSPs inappropriate for testing the
DM heating scenario.

\section{Conclusion and discussion}

We have studied the time evolution of NS surface
temperature, taking account of both the rotochemical and DM heating
effects. We have found that for ordinary pulsars the DM heating effect
can still be observed even with the rotochemical heating
if the initial period of NSs is relatively large, since in
this case the chemical imbalance does not overcome the threshold
$\Delta_{\mathrm{th}}^\infty$ and thus the rotochemical heating is
ineffective. 
The rotochemical heating operates if the initial period is as small as $\Order(1)\unit{ms}$. Thus for MSPs, the DM heating is
always concealed by the rotochemical heating. 

The surface temperature at late times depends not only on the initial
period but also on the choice of the nucleon pairing gaps, as shown in
Fig.~\ref{fig:CP_10ms}. Depending on these unknown quantities, the
rotochemical heating effect may mimic the DM heating effect in old
ordinary pulsars. For instance, the late-time temperature for the proton AO and neutron "b" gaps in Fig.~\ref{fig:CP_10ms} is kept at a few thousand K due to the rotochemical heating.
To distinguish these two heating effects, therefore,
it is necessary to improve our knowledge on nucleon pairing gaps as well
as to evaluate the initial period of pulsars accurately. We note in
passing that it is possible to estimate the initial period of a pulsar
if, for instance, the pulsar is associated with a supernova and its age
is computed from the motion of the supernova remnant, as is performed in Ref.~\cite{2012Ap&SS.341..457P},

In any case, in the presence of both the rotochemical and DM heating effects,
the late-time temperature is bounded below,
\textit{i.e.}, $T_s^\infty \gtrsim 2 \times 10^3$~K, which is determined
by the DM heating and thus independent of the initial period and pairing
gaps. As a consequence, an observation of a NS with a surface
temperature that is sufficiently below this lower bound readily excludes
the DM heating caused by typical WIMPs, and thus can severely constrain
such DM models. To see this significance, as an example, we consider the
case where DM is comprised of the neutral component of an electroweak
multiplet \cite{Cirelli:2007xd, Cirelli:2009uv, Farina:2009ez,
Farina:2013mla, Nagata:2014aoa}. This class of DM candidates includes
the pure wino/higgsino in the supersymmetric models. Although the
elastic scattering  cross section of such a
DM candidate with a nucleon is generically small \cite{Hisano:2011cs,
Hisano:2015rsa}, it can still be trapped by NSs through the inelastic
scattering with the NS matter. For this class of DM candidates, the
charged components of the DM multiplet are degenerate with the DM
component in mass, with mass differences of $\mathcal{O} (100)$~MeV,
which are smaller than the energy transfer of the DM scattering in
NSs, $\Delta E \lesssim 1$~GeV \cite{Baryakhtar:2017dbj}. As a
result, the inelastic scattering accompanied with the charged component
can occur in NSs. Since it is induced by the tree-level exchange of
the $W$ boson, its cross section is much
larger than the critical value $\sigma_{\mathrm{crit}} \simeq R^2 m_N/M
\sim 10^{-45}~\mathrm{cm}^2$ \cite{Baryakhtar:2017dbj}. Hence, we can
directly use the above results for this class of DM candidates,\footnote{There
is a small difference since the annihilation of
these DM candidates can generate neutrinos in the final state and thus
the parameter $\chi$ in Eq.~\eqref{eq:lhdm} is smaller than unity. This
difference only results in an $\mathcal{O} (1)$\% change in the
late-time temperature, which is in effect negligible in the present
discussion. } and in
particular we conclude that an observation of an old NS with $T_s^\infty
\lesssim 10^3$~K can exclude all of these DM
candidates. Notice that this constraint is independent of the DM mass as
long as it is $\lesssim 1$~PeV, which is the case for these DM
candidates, whose masses are predicted to be 1--10~TeV
\cite{Farina:2013mla}. 

Finally, we note that there are other heating mechanisms proposed in the
literature \cite{Gonzalez:2010ta}, such as the vortex creep heating
\cite{1984ApJ...276..325A, 1989ApJ...346..808S, 1991ApJ...381L..47V,
1993ApJ...408..186U, VanRiper:1994vp, Larson:1998it} and
rotationally-induced deep crustal heating \cite{Gusakov:2015kaa}. These
heating mechanisms may also compete with the rotochemical and DM heating
effects, and therefore the consequence drawn in this letter may be
altered if they are also included. We will study the implications of
these heating mechanisms for the DM heating on another occasion
\cite{HNY}.

\begin{acknowledgments}
\section*{Acknowledgments}
We thank Teruaki Enoto, Kenji Fukushima, Kazuhiro Nakazawa, Hideyuki
Umeda, and Satoshi Yamamoto for valuable discussions and suggestions.
We are also grateful to Jiaming Zheng for useful comments on the manuscript. 
This work is supported in part by the Grant-in-Aid for
Scientific Research A (No.16H02189 [KH]),
Young Scientists B (No.17K14270 [NN]), Innovative Areas
(No.19H04612 [KH], No.18H05542 [NN]).
The work of KY was supported by JSPS KAKENHI Grant Number JP18J10202.
\end{acknowledgments}

\bibliography{ref}

\end{document}